\documentclass[10pt, a4paper]{article}

\usepackage{cmap} 
\usepackage[T2A]{fontenc}

\usepackage{amssymb, amsmath}
\usepackage{cite}
\usepackage{graphicx}

\begin{document}

\title{On diagrammatic technique for nonlinear dynamical systems}

\author{
	\footnotesize Mykola Semenyakin
	\footnote{Department of Physics, Taras Shevchenko National University of Kyiv, Ukraine}
	\footnote{Bogolyubov Institute for Theoretical Physics, Ukraine}\\
	\footnotesize  semenyakinms@gmail.com
}

\maketitle

\begin{abstract}
In this paper we investigate phase flows over $\mathbb{C}^n$ and $\mathbb{R}^n$ generated by vector fields $V=\sum P^{i}\partial_i$ where $P^{i}$ are finite degree polynomials. With the convenient diagrammatic technique we get expressions for evolution operators $ev\{V|t\}: x(0)\mapsto x(t)$ through the series in powers of $x(0)$ and $t$, represented as sum over all trees of particular type. Estimates are made for the radius of convergence in some particular cases. The phase flows behavior in the neighborhood of vector field fixed points are examined. Resonance cases are considered separately.\\
\end{abstract}

\section{Introduction}

In this paper we consider transformation:
\begin{equation}
ev\{V|t\}: \mathbb{R}\times \mathbb{R}^n \to \mathbb{R}^n
\end{equation}
generated by the action of phase flow of the vector field $V$:
\begin{equation}
\lim\limits_{t\to 0}\dfrac{d}{dt}ev\{V|t\}(x_0)=V(x_0)
\end{equation}
in the case when coordinate functions of vector fields are the polynomials of finite degree. Calculation of the action of phase flows is equivalent to solving of the system of ODEs. In the study of ODEs, we often use pertrubation series methods \cite{Poincare}. In such cases we decompose right hand side of an equation into the series, take linear part, and consider all the other as small in some neighborhood:
\begin{equation}
\dot{x}=Ax+\varepsilon N(x)
\end{equation}

Here $A$ is a linear operator, $\varepsilon$ - some small parameter, $N(x)$ - some function. In cases, when it is not confusing, we will omit indexes, which mark the components of the vectors and tensors. A solution of the system is presented as series by powers of parameter $\varepsilon$:
\begin{equation}
x(t) = \sum\limits_{k=0}^{\infty}\varepsilon^k x_k(t)
\end{equation}
if it converges in some neighborhood of initial point. A term behind of $\varepsilon^0$ could be calculated from a linear differential equation:
\begin{equation}
\dot{x}=Ax
\end{equation}
whose solution is delivered by matrix exponent:
\begin{equation}
x_0(t)=\exp(At)x_0(0)
\end{equation}
\begin{equation}
ev\{A|t\}=\exp(At)=\sum\limits_{k=0}^{\infty}\dfrac{(At)^k}{k!}
\end{equation}

With linear algebra, we could calculate exponent in the basis, where basis vectors are the eigenvectors of $A$. But in case, when we want to keep nonlinear terms, such algebraic background is not developed yet \cite{IntroToNLN}.\\

An objective of this paper is to investigate one possible way to create such algebraic framework, based on the diagrams, consisting of contracted $(1,s)$-tensors. It is instructive to consider this technique on a discrete dynamic example.

\section{Discrete dynamic}

For simplicity sake, let's consider a map:
\begin{equation}
\vec{x}_{n+1}=P(\vec{x_n},...,\vec{x_n})
\end{equation}
where $P$ - is a homogeneous polynomial of degree $s$. In indexes, we can rewrite it as:
\begin{equation}
x^{i}_{n+1}=P^i_{j_1,...,j_s}x^{j_1}_n...x^{j_s}_n
\end{equation}
where a summation by the repeating indexes is introduced. The next iteration of the map can be gotten by the next iteration substituting instead of $x^{j_k}_n$:
\begin{equation}
x^{i}_{n+1}=P^i_{j_1,...,j_s}P^{j_1}_{k^1_1,...,k^1_s}x^{k^1_1}_{n-1}... x^{k^1_s}_{n-1} ... P^{j_s}_{k^s_1,...,k^s_s}x^{k^s_1}_{n-1}...x^{k^s_s}_{n-1}
\end{equation}
For convenience contraction notation, we introduce the following diagrams:\\
\pagebreak
\begin{figure}[h]
\begin{center}
\includegraphics[width=0.6\textwidth]{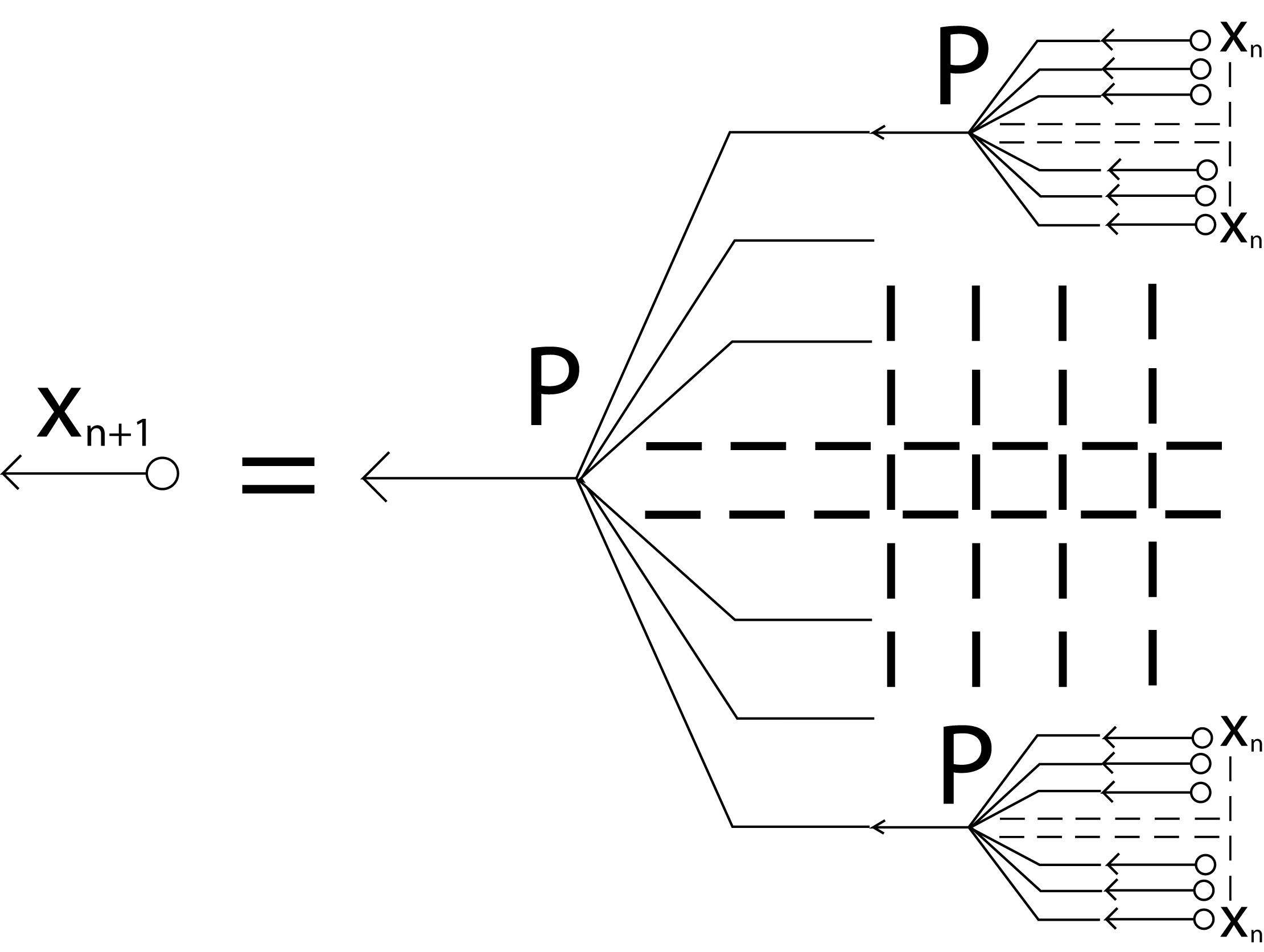}
\caption{Recurrence step as diagrams}
\end{center}
\end{figure}

where the below notations are introduced:\\
\begin{figure}[h]
\begin{center}
\includegraphics[width=0.9\textwidth]{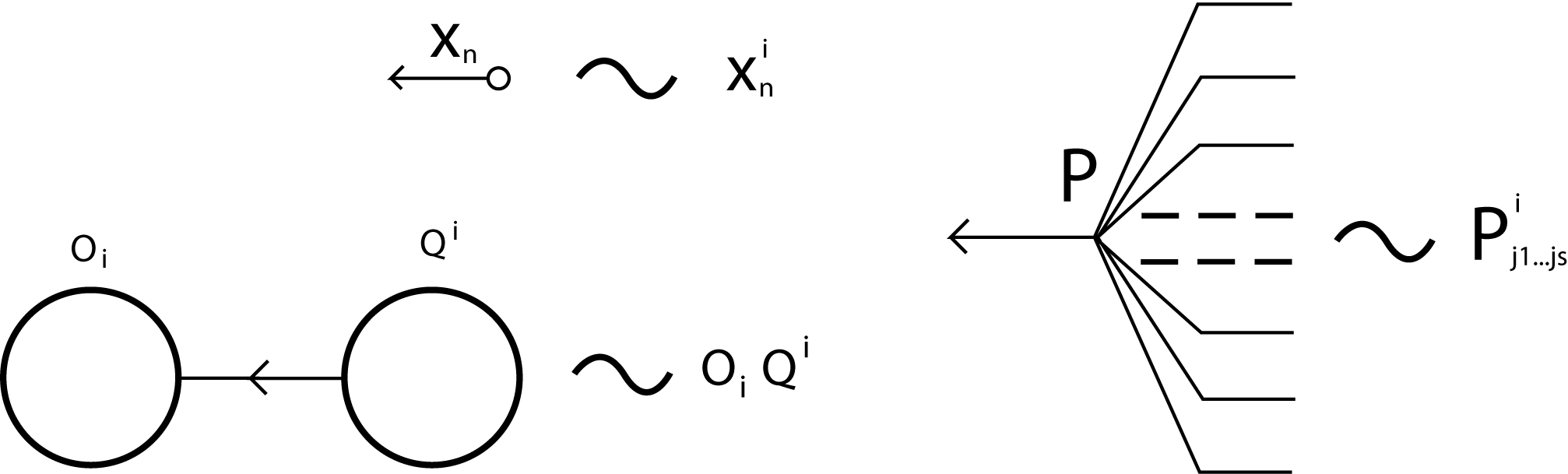}
\caption{Diagramatical notation}
\end{center}
\end{figure}

A number of operators, which diagram consists of we call \textbf{an order} of diagram, and denote $|D|$. We will omit indexes, arrows over the edges of the diagrams, and vertexes signatures in cases, when it is not confusing. Summation is to be done by the repeating indexes. The arrows direction is always from right to left.\\
For recurrence with arbitrary non-linearity on the r.h.s:
\begin{equation}
x^i_{n+1}=\sum\limits_{s=0}^{h}P^i_{j_1,...,j_s}x_n^{j_1}...x_n^{j_s}=P^i(x_n)
\end{equation}
evolution operator looks like:
\begin{equation}
x^i_n=ev^i\{P|n\}(x_0)=\sum\limits_{|D|=n}D^i(x_0)
\end{equation}
where summation goes over all possible diagrams of order $n$, which we can get by contraction of operators $P^i_{j_1,...,j_s}$. A calculation of such sum in general case is not a subject of this papers discussion.

\section{General evolution operator}

In this section and until the very end we consider general systems of differential equations with the polynomials in the right side. We will find ansatz for a generic solution and make some estimations for its converging radius. 

\subsection{Ansatz: general case}
Let's consider a system of autonomous differential equations of general form, with the polynomials in r.h.s:
\begin{equation}
\dot{x}^i=\sum\limits_{k=0}^{\infty}T^i_{j_1...j_k}x^{j_1}... x^{j_k}
\end{equation}

For simplicity, we consider homogeneous polynomials of the same degree in r.h.s. Generalization for sum of few homogeneous components is trivial, as it will become clear later.
\begin{equation}
\dot{x}^i=T^i_{j_1j_2j_3...j_s}x^{j_1}x^{j_2}... x^{j_s}
\end{equation}
\begin{equation}
\dfrac{dx}{dt}=T(x,...,x)
\end{equation}

Let's transform this equation into a recurrence. For that, let's split a time interval $t$ by $N$ similar intervals $\Delta t=\frac{t}{N}$, and rewrite equation as:
\begin{equation}
\dfrac{x_{i+1}-x_i}{\Delta t} = T(x_i,...,x_i)
\end{equation}

To get an explicit solution, we should fix $t$ and calculate limit, if it exists:
\begin{equation}
ev\{x(0)|t\}=\lim\limits_{N \to \infty}\left(E+\dfrac{t}{N}T\right)^{\circ N}x(0)
\end{equation}

Here a composition means, that with every upper index of previous composite-multiplier, we contract lover index of the next one. 

\begin{figure}[h]
\begin{center}
\includegraphics[width=0.65\textwidth]{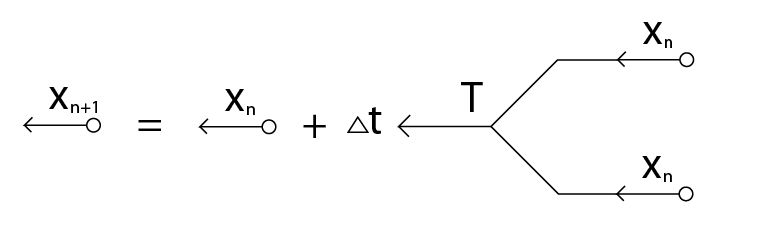}
\caption{Recurrence step for $s=2$}
\end{center}
\end{figure}

We could get the next iteration by contraction of operator $(E+\Delta t T)$ with every free edge on the right side of every diagram, and expanding by linearity. The $k^{th}$ iteration will look like:
\begin{equation}
x_k = \sum\limits^{\infty}_{n=0}(\Delta t)^n\sum\limits_{|D|=n}C^{D}_k D(x_0)= \sum\limits^{\infty}_{n=0}t^n\sum\limits_{|D|=n}\dfrac{C^{D}_k}{N^n} D(x_0)
\end{equation}
where $|D|$ - order of diagram $D$, $C^{D}_k$ - coefficient near diagram $D$ for iteration $k$, $\sum_{|D|=n}$ means summation over all diagrams of order $D$. But we need to get limit $N \to \infty$, so evolution operator looks like:
\begin{equation}
ev\{T|t\}x(0)=\lim\limits_{N \to \infty}\left(E+\dfrac{t}{N}T\right)^Nx(0)=\sum\limits^{\infty}_{n=0}t^n\sum\limits_{|D|=n}C^{D}_{\infty} D(x(0))
\end{equation}
where $C^{D}_{\infty} = \lim\limits_{N\to\infty}\dfrac{C^D_{N}}{N^{|D|}}$ - asymptotic coefficient. We could get its value by direct calculation, but we will use some trick here. For r.h.s. with the several homogeneous components an answer will have the same form, but in the diagrams there will also be forks, corresponding to all homogeneous components of r.h.s. (in the further text, operators in diagrams would be sometimes referred as the 'vertices').

\subsection{Calculation of the asymptotic coefficients}

The trick is that we substitute a solution:
\begin{equation}
x(t)=\lim\limits_{N\to\infty}x_N=\sum\limits^{\infty}_{n=0}t^n\sum\limits_{|D|=n}C^{D}_{\infty} D(x(0))
\end{equation}
in the equation ($T$ - operator of order $s$):
\begin{equation}
\dfrac{dx}{dt}=T(x,...,x)
\end{equation}
\begin{equation}
\sum\limits_{D}(|D|+1)t^{|D|}C^{D}_{\infty}D^i(x(0))=T^i_{j_1...j_s} \sum\limits_{D_1} t^{|D_1|}C^{D_1}_{\infty} D_1^{j_1}(x(0))...\sum\limits_{D_s}t^{|D_s|}C^{D_s}_{\infty} {D_s}^{j_s}(x(0))
\end{equation}

Now, if we compare the coefficients before the same diagrams, we get:
\begin{equation}
C^{D}_{\infty}=\dfrac{C^{D_1}_{\infty}C^{D_2}_{\infty}...C^{D_n}_{\infty}}{|D|}
\end{equation}
where the diagram $D$ arises from the operator $T$, to the free edges of which we have contracted $D_1$,$D_2$...$D_n$:

\begin{figure}[h]
\begin{center}
\includegraphics[width=0.5\textwidth]{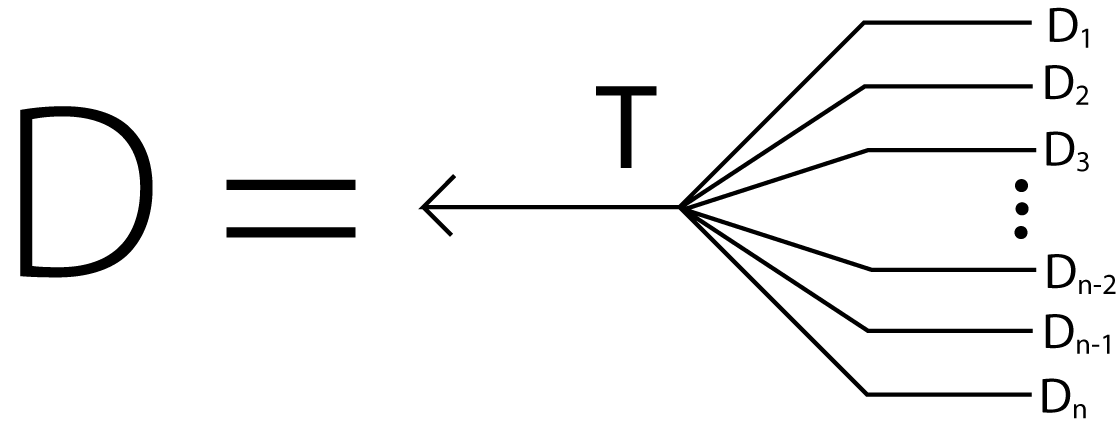} 
\caption{Diagram $D$ in recurrence for $C^{D}_{\infty}$}
\end{center}
\end{figure}

In this formula, we could substitute $C^{D_1}_{\infty},C^{D_2}_{\infty},...,C^{D_n}_{\infty}$ in the same way, and so on. We repeat this process, until we come to the trivial diagram, which consists of one identical operator, a coefficient before which is 1. Thus, the coefficients look like: 
\begin{equation}
\label{coefficient_general}
\dfrac{1}{C^{D}_{\infty}}=\prod\limits_{K\preceq D}|K|=D!
\end{equation}
where product is calculated by the sub-diagrams, which we could get by taking out all sub-trees, which 'originate' from the each vertex, and have the same right, 'end', edges as $D$. Generally:
\begin{equation}
A\preceq B \Leftrightarrow \exists C, q: B^{i}_{j_1...j_s}=C^i_{j_1...j_q...j_p}A^{j_q}_{j_{p+1}...j_s}
\end{equation}
where $C$ - some diagram, which is also constructed from the basic vertexes. We will use the same notation in the further text. In such way, a general form of evolution operator is:
\begin{equation}
\boxed{
$$
\label{general_exponent}
ev\{T|t\}x(0)=\sum\limits_{D}\dfrac{t^{|D|}}{D!}D^i(x(0))$$
}
\end{equation}
which really looks like a matrix exponent in which we sum up by all 'branching' degrees of the diagrams. It is the key formula of this text.

\subsection{Convergence of evolution operators expansion }

The folloiwng inequality holds for Euclidean norm of contraction of two operators: 
\begin{equation}
\parallel A^i_{a_1...a_p}B^{a_1}_{b_1...b_p} \parallel < \parallel A \parallel \parallel B\parallel
\end{equation}

Thus, for evolution operator for equation with homogeneous non-linearity of degree $s$ we could make an estimation:
$$
|x(t)|=\left|\sum\limits_{p=0}^{\infty}t^p\sum\limits_{|D|=p}C^D_\infty D(x_0)\right| < \sum\limits_{p=0}^{\infty}t^p\sum\limits_{|D|=p}C^D_\infty \parallel T \parallel ^p |x|^{p(s-1)+1}=
$$
\begin{equation}
=\sum\limits_{p=0}^{\infty} C^p_\Sigma \parallel T \parallel ^p t^p |x|^{p(s-1)+1}
\end{equation}
where $C^p_\Sigma = \sum\limits_{|D|=p}C_\infty^D$. So, we can also estimate a convergence radius of the series:
\begin{equation}
\label{convergence_estimation_1}
t|x|^{s-1}<\lim\limits_{p\to\infty}\dfrac{1}{\sqrt[p]{C^p_\Sigma \parallel D \parallel^p }} = \dfrac{1}{\parallel D \parallel }\lim\limits_{p\to\infty}\dfrac{1}{\sqrt[p]{C^p_\Sigma}}
\end{equation}

We get coefficients $C^p_\Sigma$ from the one dimensional equation of such kind:
\begin{equation}
\dot{x} = \alpha x^s
\end{equation}
\begin{equation}
x(0)=x_0
\end{equation}
After integration, we get:
\begin{equation}
-\dfrac{s-1}{x^{s-1}}=\alpha t+c
\end{equation}
\begin{equation}
x(t)=\dfrac{x_0}{\sqrt[s-1]{1-\dfrac{\alpha x_0^{s-1}}{s-1}t}}
\end{equation}
\begin{equation}
\label{solution_1}
x(t)=\sum\limits^{\infty}_{k=0}\dfrac{(\alpha t)^kx_0^{k(s-1)+1}}{(s-1)^k(k-1)!}\prod\limits^{k-1}_{p=0}(p+\dfrac{1}{s-1}) = \sum\limits^{\infty}_{k=0}\dfrac{(\alpha t)^kx_0^{k(s-1)+1}}{(s-1)^k}\dfrac{\Gamma((s-1)^{-1}+k)}{\Gamma(k)\Gamma((s-1)^{-1})}
\end{equation}
Space is one dimensional, therefore:
\begin{equation}
T(x,x...,x)=T^0_{00...0}x^s=\alpha x^s
\end{equation}

So, all diagrams with the same number of the vertices will have same values. From solution (\ref{solution_1}) we could get the coefficients:
\begin{equation}
\label{c_sigma_k}
C^k_{\Sigma}=\sum\limits_{|D|=k}C^{D}_{\infty}=\dfrac{1}{(1-s)^k}\dfrac{\Gamma((s-1)^{-1}+k)}{\Gamma(k)\Gamma((s-1)^{-1}} = \dfrac{1}{(1-s)^k}B((s-1)^{-1},k)^{-1}
\end{equation}
After substitution of (\ref{c_sigma_k}) into (\ref{convergence_estimation_1}), we get
\begin{equation}
t|x|^{s-1} < \dfrac{s-1}{\parallel D \parallel }
\end{equation}

In the same way we could get estimation for convergence radius for more complicated r.h.s., which consist of more quantity of the homogeneous components. Difficulty here is, that during the process of calculating of $C^k_{\Sigma}$ we need to solve algebraic equations of higher degree, to calculate $x(t,x_0)$ in the one dimensional case.

\section{Perturbations theory: non-resonance case}

In this section we consider how, if we know solution for a linear system, to built non-linear perturbations theory over it. We consider  ansatz substituting procedure as explained in the previous section. Estimations for convergence radius for a non-resonance cases are also presented. As to classical results on dynamic in the neighbourhood of the fixed points see, for example,\cite{Siegel} or \cite{Grobman}.

\subsection{Ansatz: 'interaction representation'}
If we decompose r.h.s. of equation near the not degenerated fixed point, equation gets the form:
\begin{equation}
\dot{x}=Ax+P(x)
\end{equation}
where $A$ - some matrix, $P$ - finite degree polynomial. Let's do substitution, which is called in physics an 'interaction representation':
\begin{equation}
x(t)=e^{At}y(t)
\end{equation}
\begin{equation}
\dfrac{d}{dt}e^{At}y(t)=Ae^{At}y(t)+e^{At}\dot{y}(t)=Ae^{At}y(t)+P(e^{At}y(t))
\end{equation}
\begin{equation}
e^{At}\dot{y}(t)=P(e^{At}y(t))
\end{equation}
\begin{equation}
\dot{y}=e^{-At}P(e^{At}y)
\end{equation}

Now, let's rewrite it as an integral equation, and iterate: 
\begin{equation}
y(t)=x_0+\int\limits_{0}^{t}dt_1e^{-At_1}P(e^{At_1}y(t_1))
\end{equation}
\begin{equation}
y(t)=x_0+\int\limits_{0}^{t}dt_1\int\limits_{0}^{t_1}dt_2e^{-At_1}P(e^{At_1}x_0+e^{At_1}e^{-At_2}P(e^{At_2}y(t_2)))
\end{equation}
and so on. Here, all calculations are on the level of formal series, a convergence question will be discussed later.\\

So, a solution looks like a sum by all diagrams, where the vertexes connected with exponents-propagators, are orderly integrated by all intermediate times. In case, when a linear operator is diagonalizable, the propagators looks like:
\begin{equation}
D^{ij}(t)=e^{(\lambda_j-\lambda_i)t}
\end{equation}
where $\lambda_i$ - eigenvalues of matrix $A$. Specific of the these series, in compare with the previous one, is that here the indexes 'inside' of the diagram play an important role - eigenvalues in intermediate exponents depend on them. So, if we want to rewrite answer as a sum over all diagrams, we should sum up by all 'skeleton' diagrams - every vertex corresponds to one component of nonlinear operator $T^i_{j_1...j_s}$ only (for example, $T^1_{23}T^2_{45}$ and $T^1_{23}T^3_{45}$ calculated separately). Thus, ansatz looks like:
\begin{equation}
x(t)=\sum\limits_{\tilde{D}}\sum\limits_{\vec{n}\in\mathbb{Z}^d}C(\tilde{D},\vec{n})e^{\vec{n}\vec{\lambda}t}\tilde{D}(x(0))
\end{equation}
where $C(\tilde{D},\vec{n})$ - some coefficients, depending on $\lambda$, $\vec{\lambda}=(\lambda_1,...,\lambda_d)$ - vector, which consists of eigenvalues, $\vec{n}=(n_1,...,n_d) \in \mathbb{Z}^d$. In case, when operator is not diagonisable, or exist resonances ($\exists \vec{n}: (\vec{\lambda},\vec{n})=0$), terms like $At^ke^{\alpha t}$ would also arise. 

\subsection{Calculation of the coefficients: non-resonance case}

Let's consider, for simplicity, an equation in $d$-dimentional space:
\begin{equation}
\label{equation_non_resonance_case}
\dot{x}^i=\lambda^i x^i+T^i_{j_1...j_s}x^{j_1}...x^{j_s}
\end{equation}
Generalization is trivial again. Let's look for its solution in the following form:
\begin{equation}
\label{ansatz_interaction_rep}
x^i(t)=\sum\limits_{D}\sum\limits_{\vec{n}\in\mathbb{Z}^d}C_D^i(\vec{n},\lambda)e^{\vec{n}\vec{\lambda}t}D^i
\end{equation}
where summation goes by all diagrams, which consist of all monomial operators, e.g. different terms in summation in contraction goes with the different coefficients $C_D^i(\vec{n},\lambda)$, $D$ - skeleton diagram. We suppose, that frequencies is not resonance: $\forall \vec{n}\in\mathbb{Z}^d \quad (\vec{\lambda},\vec{n}) \neq 0$. Here it is important, that on right side of the diagrams not $x^i(0)$ is contracted, but some coefficients $c_0^i$, which are not known, but depended on initial conditions. Substituting (\ref{ansatz_interaction_rep}) in (\ref{equation_non_resonance_case}), we get:
$$
\sum\limits_{P}\sum\limits_{\vec{p}\in\mathbb{Z}^d}\vec{p}\vec{\lambda} C_P^i(\vec{p})e^{\vec{p}\vec{\lambda}t}P^i=
\sum\limits_{Q}\sum\limits_{\vec{q}\in\mathbb{Z}^d}\vec{e}_i\vec{\lambda} C_Q^i(\vec{q})e^{\vec{q}\vec{\lambda}t}Q^i+
$$
\begin{equation}
\sum\limits_{\{D_i\}}\sum\limits_{\{\vec{d}_i\}\in\mathbb{Z}^d} T^i_{j_1...j_s}D_1^{j_1}...D_s^{j_s} C^{j_1}_{D_1}(\vec{d}_1)...C^{j_s}_{D_s}(\vec{d}_s)e^{(\vec{d}_1+...+\vec{d}_s)\vec{\lambda}t}
\end{equation}
where $\vec{e}_i=(0,...,1,...,0)$ - unit vector along $i$-th axis. The coefficients before the similar exponents and diagrams are:
\begin{equation}
\vec{p}\vec{\lambda} C_P^i(\vec{p})=\vec{e}_i\vec{\lambda} C_P^i(\vec{p}) +\sum\limits_{\vec{d}_1+...+\vec{d}_s=\vec{p}} C^{j_1}_{D_1}(\vec{d}_1)...C^{j_s}_{D_s}(\vec{d}_s)
\end{equation}
\begin{equation}
(\vec{p}-\vec{e}_i)\vec{\lambda} C_P^i(\vec{p})=\sum\limits_{\vec{d}_1+...+\vec{d}_s=\vec{p}} C^{j_1}_{D_1}(\vec{d}_1)...C^{j_s}_{D_s}(\vec{d}_s)
\end{equation}
\begin{equation}
C_P^i(\vec{p})=\dfrac{1}{(\vec{p}-\vec{e}_i)\vec{\lambda}}\sum\limits_{\vec{d}_1+...+\vec{d}_s=\vec{p}} C^{j_1}_{D_1}(\vec{d}_1)...C^{j_s}_{D_s}(\vec{d}_s)
\end{equation}

Now, we can substitute every $C^{j_a}_{D_a}$ for its decomposition, and repeat this process until the diagrams without vertices remain. A coefficient before a term, without vertices is:
\begin{equation}
\sum\limits_{\vec{p}\in\mathbb{Z}^d}\vec{p}\vec{\lambda} C_0^i(\vec{p})e^{\vec{p}\vec{\lambda}t}=
\sum\limits_{\vec{q}\in\mathbb{Z}^d}\vec{e}_i\vec{\lambda} C_0^i(\vec{q})e^{\vec{q}\vec{\lambda}t}
\end{equation}
\begin{equation}
\sum\limits_{\vec{p}\in\mathbb{Z}^d}(\vec{p}-\vec{e}_i)\vec{\lambda} C_0^i(\vec{p})e^{\vec{p}\vec{\lambda}t}=0
\end{equation}

Thus, only non-zero coefficient can be put before the term, where $\vec{p}=\vec{e}_i$. Now, coming back along the diagram from its right side to the left one, we see that in every vertex only one exponent is left, which is defined by the indexes on its right edges.
\begin{equation}
x(t)=\sum\limits_{D}C^{i}_DD^i_{j_1...j_r}c_0^{j_1}...c_0^{j_r}e^{(\vec{e}_{j_1}+...+\vec{e}_{j_r})\vec{\lambda}t}
\end{equation}
\begin{equation}
C^i_D=\prod\limits_{v\in D}(-\lambda_{i(v)}+\sum\limits_{k\in v\to^{All}}\lambda_{i(k)})^{-1}
\end{equation}
where product is done by all diagrams vertexes, and a nested sum - by all free edges, which are on the right side from vertex $v$. Symbol $i(k)$ means index, which corresponds to edge, $i(v)$ - index, which corresponds to the left edge, which enters to diagram. For the further convenience, we could rewrite it as:
\begin{equation}
\label{coefficient_resonance}
C^i_D=\prod\limits_{K\preceq D}\left(\sum\limits_{v\in K} \lambda(v)\right)^{-1}= \prod\limits_{K\preceq D}\left(\lambda(K)\right)^{-1}
\end{equation}
where product is calculated by the sub-diagrams, which we could get by taking out all sub-trees, which 'originate' from the each vertexes (as in (\ref{coefficient_general})), and  $\lambda(v)$ means a sum of all eigenvalues, which correspond to the edges, which originate from the vertex (or diagram), with the correct sign:
\begin{equation}
\label{definition_lambda}
\lambda(P^{i_1...i_u}_{j_1...j_d})=\lambda_{j_1}+\lambda_{j_d}-\lambda_{i_1}-...-\lambda_{i_u}
\end{equation}
This function is 'good' according to the contraction (denoted with $\star$):
\begin{equation}
\lambda(P\star Q)=\lambda(P)+\lambda(Q)
\end{equation}
The coefficients $c_0^{i}$ are defined from initial conditions through equation:
\begin{equation}
x^i(0)=x_0^i=\sum\limits_{D}C_D^i D^i_{j_1...j_r}c_0^{j_1}...c_0^{j_r}
\end{equation}
So, dynamic in the neighborhood of a fixed point looks like:
\begin{equation}
x(t)=f(c_0^1 e^{\lambda_1 t},c_0^2 e^{\lambda_2 t},...,c_0^d e^{\lambda_d t})
\end{equation}
where $f$ is the function, whose coefficients we just calculated, and the variables $c_0^i$ are
multiplied by the corresponding exponents. The constants dependency on initial conditions can be expressed as:
\begin{equation}
x(t)=f\circ e^{At}\circ f^{-1}(x_0)
\end{equation}
where $f^{-1}$ - a function inverse to $f$. Its explicit form can be obtained, if we suppose, that it could be represented as the series by diagrams, with some coefficients:
\begin{equation}
f^i(x)=\sum\limits_D C_D D^i_{j_1...j_n}x^{j_1}...x^{j_n}
\end{equation}
\begin{equation}
(f^{-1})^i(x)=\sum\limits_D G_D D^i_{j_1...j_n}x^{j_1}...x^{j_n}
\end{equation}
\begin{equation}
f^{-1}\circ f = id
\end{equation}
If we do same procedure with $f^{-1}\circ f = id$, as in the previous cases, we get:
\begin{equation}
G_D=\sum_{\pi\in\mathcal{P}(D)}(-1)^{\#(\pi)+1}\prod\limits_{Q\in\pi}C_Q
\end{equation}
where $\mathcal{P}(D)$ - all possible ways to split diagram on sub-diagrams over edges, $\#(\mathcal{P}(D))$ - number of the sub-diagrams in partition, $\prod_{Q\in\pi}$ - product by all sub-diagrams in partition.

\subsection{Perturbation series convergence}

Here we clearly see a 'small denominators' problem - even in non-resonance case, denominators $(\vec{\lambda},\vec{n})$ can be arbitrary small. From Diophantine approximations theory \cite{SmallDenominators}, we know, that for the majority of $\vec{\lambda}$ (except Lebesgue measure zero) the following estimation is applicable:
\begin{equation}
\exists M \in \mathbb{R} \forall \vec{n}\in\mathbb{Z}^d:|(\vec{\lambda},\vec{n})| > M|n|^{-\nu},\nu=d+1
\end{equation}

For simplicity, let's consider a case, when diagrams consist only of operators with $s$-forks. Asymptotically:
$$
\prod\limits_{v\in D}1/|-\lambda_{i(v)}+\sum\limits_{k\in v\to^{All}}\lambda_{i(k)}| < \prod\limits_{v\in D}M^{-1}|-\vec{e}_{i(v)}+\sum\limits_{k\in v\to^{All}}\vec{e}_{i(k)}|^{\nu} \sim 
$$
\begin{equation}
M^{-|D|}(s-1)^{|D|\nu} (D!)^{\nu}
\end{equation}
$$|f(x)|<\left| \sum\limits_D  D^i_{j_1...j_n}x^{j_1}...x^{j_n} M^{-|D|}(s-1)^{|D|\nu} (D!)^{\nu} \right| <$$
$$\sum\limits_D \parallel T \parallel^{|D|} |x|^{(s-1)|D|} M^{-|D|}(s-1)^{|D|\nu} (D!)^{\nu}=$$
\begin{equation}
\sum\limits_{n=0}^{\infty}  \left(\dfrac{\parallel T \parallel |x|^{s-1}(s-1)^{\nu}}{M}\right)^n \sum\limits_{|D|=n}(D!)^{\nu}
\end{equation}
Estimation for convergence radius:
\begin{equation}
|x|^{s-1}<\dfrac{M}{(s-1)^{\nu} \parallel T \parallel}\left(\lim\limits_{n\to\infty}\sqrt[n]{\sum\limits_{|D|=n}(D!)^{\nu}}\right)^{-1}
\end{equation}
In the same way, we could get one for $f^{-1}$:
\begin{equation}
|x|^{s-1} < \dfrac{M}{(s-1)^{\nu} \parallel T \parallel} \left(\lim\limits_{n\to\infty}\sqrt[n]{\sum\limits_{|D|=n} \sum_{\pi\in\mathcal{P}(D)}(-1)^{\#(\pi)+1}\prod\limits_{Q\in\pi}(Q!)^{\nu} } \right)^{-1}
\end{equation}

\section{Perturbations theory: resonance case}

In this section, we will consider perturbations theory in the case of presence of resonances. To calculate series for solution, we will develop new approach, based on the direct perturbation series coefficients calculation.

\subsection{Ansatz: ordered exponent}

Here we come up to the calculation of the evolution operator directly from the limit transition. We represent $(E+\Delta t P)^{\circ N}$ 'decomposed' along the time - when we calculate composition, we assign composition multiplier to the time, when it acts, like an infinitesimal transformation, and consider the terms in particular order:

\begin{figure}[h]
\begin{center}
\includegraphics[width=1\textwidth]{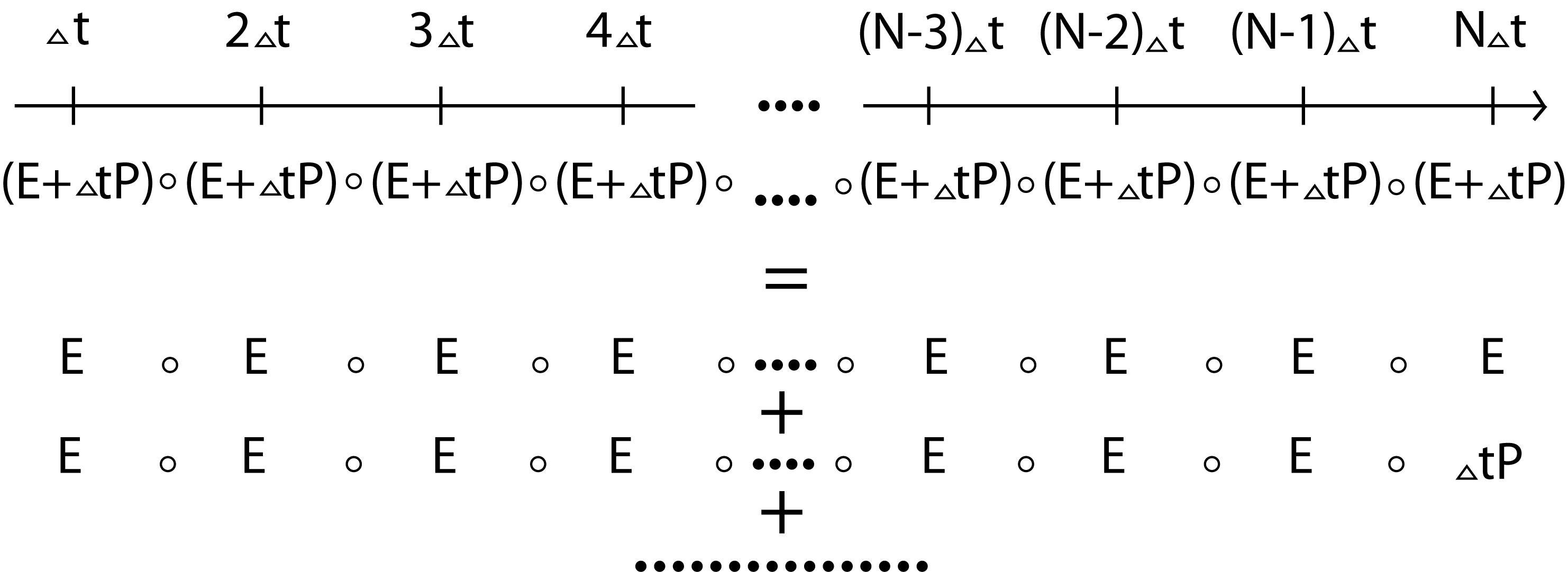}
\caption{Evolution operator ordered in time}
\end{center}
\end{figure}

Let's consider an equation:
\begin{equation}
\dot{x}=Ax+P(x)
\end{equation}
where $P(x)$ - again, for simplicity, homogeneous operator - generalization for larger number of homogeneous components is obvious. Evolution operator looks like:
\begin{equation}
ev\{A,P|t\}=\lim\limits_{N\to\infty}(E+\Delta t A+\Delta t P)^{\circ N}
\end{equation}

Let's suppose, we have expanded all brackets. Let's consider the terms in order of increasing of number of occurrence of $P$ in them. The first term doesn't have $P$ in it at all. Calculating composition from left to the right, we get:
\begin{equation}
P^0: \lim\limits_{N\to\infty} (E+\Delta t A)^{\circ N}=ev\{A|t\}=e^{At}
\end{equation}

Now, let's consider the terms, with one occurrence of $P$ in them. Let's consider all terms, were $P$ appear on $k$-th position. Thus, summation by all terms with different number of appearing of $A$ gives us the ordinal exponents before, and after $k$-th position:
\begin{equation}
P^1: \lim\limits_{N\to\infty}\sum\limits_{k=0}^N e^{Ak\Delta t}\circ P \circ e^{A(N-k)\Delta t}\Delta t
\end{equation}
where the second composition is got on all edges. In a limit transition $N\to\infty$, a sum by $k$ should be replaced by the below integral:
\begin{equation}
\label{example_ordered_exp}
(P^1)^i_{mn}: \int\limits_0^t dt_1 (e^{A(t-t_1)})^i_j P^j_{kl} (e^{At_1})^k_m (e^{At_1})^l_n
\end{equation}
\begin{equation}
(P^2)^i_{rsf}: \int\limits_0^t dt_2\int\limits_{0}^{t_2} dt_1 (e^{A (t-t_2)})^i_j P^j_{kl}  (e^{A(t_2-t_1})^k_m P^m_{pq} (e^{At_1})^p_r (e^{At_1})^q_s (e^{A t_2})^l_f
\end{equation}
$$.....$$
and so on. Here we should be careful with a times order $0<t_1<t_2<t$.

\subsection{Resonances}

Here, we consider separately all terms of non-linearity, assuming that linear part is diagonalised, without Jordan cells, e.g.:
\begin{equation}
\label{example_resonances}
T^i_{jk}=T^x_{xx}+T^{y}_{xy}+T^{x}_{yz}
\end{equation}

Since we aren't interested in the particular coefficients values, but in existence of the terms with some particular indexes only, we represent (\ref{example_resonances}) in the form:
\begin{equation}
T\sim x\to {xx}+y\to xy+x\to yz
\end{equation}
Contraction of indices gives: 
\begin{equation}
T^x_{yz}T^y_{xx}\sim x\to yz\circ y\to xx=x\to zxx
\end{equation}

Using such approach, we could look at the diagrams in evolution operator from a little bit different point of view - it has a sum of all possible terms, generated by such substitutions inside of it. Existence of substitution in the set of generators implies an existence of corresponding component in the r.h.s. of equation. With every such a diagram we have an ordered integration by time. Thus, we will have the terms like:
\begin{equation}
D \sim x\to yyzzzz \int\limits_{t_1}^{t_2}dt e^{t(2\lambda_y+4\lambda_z-\lambda_x)}
\end{equation}

If a factor multiplied by $t$ in the exponent is zero, a term with $t^n$ is present in solution. But this relates not to such diagram only, but to all diagrams, which have such sub-diagram. In some specific cases, it could be reduced, but this is not the case for the general coefficients.\\

A criterion of existence of resonance terms is existence of solution in natural numbers (or with one $-1$ - corresponding to the left edge) of equation:  
\begin{equation}
\lambda_1 n_1+\lambda_2 n_2+...+\lambda_d n_d=0
\end{equation}

Even in case of three $\mathbb{R}$-valued eigenvalues it is a non-trivial question. But let us imagine, that we have solved this equation. The next necessary question is if such term would be delivered by generators. Necessary condition is existence of solution of matrix equation in $\mathbb{N}$:
\begin{equation}
A\vec{v}=\vec{n}
\end{equation}
where $A$ - matrix, which columns are the generators, with multiplicity of every index (e.g. $x$ or $y$) in every cell, $\vec{n}$ - solution, for resonance set of eigenvalues. But even if this equation has a solution, still there is an open question, if we are really able to build such diagram from the generators.\\

In terms of function $\lambda$ defined in the (\ref{definition_lambda}), every resonance is corresponding to the solution of the equation:
\begin{equation}
\lambda(D)=0
\end{equation}
for some D, generated by r.h.s of equation.\\
\textbf{Example.} Let's consider two-dimensional system, with commensurable eigenvalues $\lambda_1=m\lambda$, $\lambda_2=-n\lambda$, $n\lambda_1 + m\lambda_2=0$. Let's assume, for example, that in a diagonal basis, its non-linearity has the terms $x\to x^a y^b$, $y\to x^c y^d$. Thus, a criterion of existence of resonance terms is:
\begin{equation}
\left( \begin{array}{cc}
a & c\\
b & d\\
\end{array} \right)
\left( \begin{array}{c}
v_x\\
v_y\\
\end{array} \right)
=
\left( \begin{array}{cc}
nk\\
mk\\
\end{array} \right)
\end{equation}
for some positive $k, v_x, v_y$. Such Diophantine systems are well investigated in the numbers theory \cite{NumbersTheory}.\\
\textbf{Example.} Numerical simulation\\
Verification was done for systems in $\mathbb{C}^2$ with the pure imagine eigenvalues $\dfrac{i\omega_1}{i\omega_2}=k$. The results are presented in an appendix.\\

The existence or absence of resonance terms could change perturbations theory completely. Ansatz for system with resonance terms would look like:
\begin{equation}
x(t)=\sum\limits_{D}\sum\limits_{\vec{n}\in\mathbb{Z}^d}\sum\limits_{k=0}^{\infty}C_D(k,\vec{n})t^ke^{\vec{n}\vec{\lambda}t}D(x(0))
\end{equation}

Getting of coefficients from the direct ansatz substitution is technically very hard, so we come to resonances problem from the ordered integration point of view.

\subsection{Calculation of the coefficients: resonance case}

For the beginning, let's calculate an integration on one vertex - something like (\ref{example_ordered_exp}). We could rewrite it in the diagonal basis and regroup eigenvalues behind coinciding times:
$$\int\limits_0^{t_2} dt_1 (e^{A(t_2-t_1)})^i_j P^j_{kl} (e^{At_1})^k_m (e^{At_1})^l_n = \int\limits_0^{t_2} dt_1  P^i_{mn} e^{\lambda_i t_2} e^{t_1 (\lambda_m+\lambda_n-\lambda_i)}=$$
\begin{equation}
\label{ordered_exponent_one_vertex}
P^i_{mn}\dfrac{e^{(\lambda_m+\lambda_n)t_2}-e^{\lambda_i t_2}}{\lambda_m+\lambda_n-\lambda_i}
\end{equation}

If this integration is the step in the calculations in the bigger diagram, then we subdivide further calculations onto two scenarios:
\begin{enumerate}
\item We take the first term in the sum. Eigenvalues are accumulating - it means, that when we will calculate next diagram, which action is located in time $t_2$, then eigenvalue $\lambda_i$ will be canceled, and we could think about edge $i$ as it was replaced with edges $m$ and $n$, with the corresponding exponents.
\item We take the second term in the sum. Eigenvalues accumulation is 'interrupted', which means, that when we will calculate next diagram, which action is located in time $t_2$, then eigenvalue $\lambda_i$ will not be canceled, and all previous part of diagram is 'died out' - indexes $m$ and $n$ will not take part in the further calculations, and we need to multiply the diagram on $-1$. We will interpret further edge $i$ as the 'beginning' edge with $\lambda_i$ in exponent only.
\end{enumerate}

And the same algorithm is applied for the all other vertexes. Now, we are ready to write down an answer for a non-resonance case:
\begin{enumerate}
\item The first summation goes over all diagrams $D^{i}_{j_1...j_s}$, which could be constructed from the r.h.s. of equation.
\item The second summation goes over all partitions $\pi\in\mathcal{P}(D)$ of diagrams. Each edge, where diagram is splitted into two sub-diagrams, corresponds to taking of the second term in the (\ref{ordered_exponent_one_vertex}), and each solid sub-diagram corresponds to the first terms, to the integration without 'interruptions'.
\item In each term in the nested summations we have $D^i_{j_1...j_s}x^{j_1}(0)...x^{j_s}(0)$.
\item Each sub-diagram $S\in\pi$ is multiplied on the $(-1)$ and on the coefficient: $\left(\prod_{K\preceq S}\sum_{v\in K} \lambda(v) \right)^{-1}$, which comes from the integration on the each vertex of the 'solid' sub-diagram. The whole diagram is multiplied by the additional $(-1)$. General degree of $(-1)$ behind each diagram is the same, as the number of the sub-diagrams in the partition, plus $1$.
\item Let's $F^i_{j_1...j_f}\in \pi$ be the only sub-diagram, which left edge is not contracted. The diagram $D$ is multiplied on the $e^{\vec{\lambda}(\vec{e}_{j_1}+...+\vec{e}_{j_f})t}$, corresponding to the right edges of the sub-diagram, which had not died out until the very end of the integration along diagram $D$.
\end{enumerate}

In general, that all gives us a result coinciding with the one, obtained from the direct ansatz substitution. We could introduce a 'normalized' diagram, to simplify the answer:
\begin{equation}
\tilde{D}^i_{j_1...j_s}=(-1)D^i_{j_1...j_s}\left( \prod_{S\preceq D}\sum_{v\in S} \lambda(v)\right)^{-1}=(-1)D^i_{j_1...j_s}C_{D}
\end{equation}
\begin{equation}
x^{i}(t)=(-1)\sum\limits_{\tilde{D}}\tilde{D}^i_{j_1...j_s} e^{\vec{\lambda}(\vec{e}_{j_1}+...+\vec{e}_{j_s})t}c^{j_1}...c^{j_s}
\end{equation}
\begin{equation}
c^{j_p}=\sum\limits_{Q}\sum\limits_{\pi\in\mathcal{P}(Q)}{\tilde{Q_\pi}}^{j_p}_{k_1...k_q}x^{k_1}(0)...x^{k_q}(0)
\end{equation}
where $C_{S}$ has been defined in (\ref{coefficient_resonance}), and $\tilde{Q_\pi}$ means, that we took diagram's partition, 'normilize' each sub-diagram, and contract 'normilized' sub-diagrams in the same order, as in the original diagram. \\

In the case of presence of resonances, we have got much more interesting picture. In the vertex, were resonance was 'born', we have simply:
\begin{equation}
\int\limits_{0}^{t_1}dt_2 1=t_2
\end{equation}
(as it was mentioned in the previous subsection). We could consider this moment like a specific 'interrupting' in the meaning, that exponent accumulation is interrupted, or as a 'birth' of the resonance. In every diagram there could be several such moments. $t_2$, which has been born in the resonance will be preserved in the further integration, if we do not choose 'dying out' alternative. Such 'resoncnce lines' (integration with $t$, which getting start on the some resonance diagram's 'initial' vertex) could meet each other, and degree of $t$ would increase. Integration on a vertex with the arbitrary degree of $t$ will look like:
$$
\int\limits_{0}^{t_1}dt_2 t_2^{s}e^{\lambda t_2}=\dfrac{t_1^{s}}{\lambda}e^{\lambda t_1} - \dfrac{st_1^{s-1}}{\lambda^2}e^{\lambda t_1} + \dfrac{s(s-1)t_1^{s-2}}{\lambda^3}e^{\lambda t_1}-...+\dfrac{(-1)^{s}s!}{\lambda^{s+1}}e^{\lambda t_1} - \dfrac{(-1)^{s}s!}{\lambda^{s+1}}=
$$
\begin{equation}
\dfrac{e^{\lambda t_1}}{\lambda}\left(t_1^{s} + \dfrac{st_1^{s-1}}{-\lambda} + \dfrac{s(s-1)t_1^{s-2}}{(-\lambda)^2}+...+\dfrac{s!}{(-\lambda)^s}\right) + \dfrac{1}{-\lambda}\dfrac{s!}{(-\lambda)^{s}}
\end{equation}
where $e^{\lambda t_1}/\lambda$ - ordinal thing, which we have seen in the previous calculations. So, we could also use 'normalization', but with the small changes - if the sub-diagram $S$ is resonancial, we shouldn't divide it on the last coefficient $\lambda(S)$ - because it is zero. To interpret terms in the brackets we should consider each 'resonance line' which comes to this vertex separately, like a new degree of freedom, and look to each $t$-degree decreasing like a 'die' of one of these 'resonance line'. As in the non-resonance case, in case of dying out of each 'resonance line' we should multiply expression by the $(-1)/\lambda$. Factor $s!/(s-k)!$ behind $t^{s-k}$ means the numbers of ways, how we could chose $k$ 'resonance lines' from $s$ for the 'dying out'. So, in some meaning, 'overlaped' 'resonance lines' behaves 'independently' - every 'resonance line' in the integration behave, like an independent one, and their evolution 'scenarios' are 'multiplied'. Each resonance line in diagram $D$, which takes part in the integration between two resonance vertexes $v_1$ and $v_2$, and 'die' between them, will give a multiplier:
\begin{equation}
L_D(v_1,v_2)=-\sum\limits_{v\in [v_1, v_2]}\lambda_v^{-1}
\end{equation}
where $[v_1, v_2]$ - set of all vertexes, lying between $v_1$ and $v_2$, $\lambda_v$ - eigenvalues sum, corresponding to the vertex $v$.

Now a case of 'nested' resonance diagrams is left unclear . Let $D$ be some diagram, $S\preceq D$ - some resonance sub-diagram: $\lambda(S)=0$. We call $Q$ 'nested' resonance sub-diagram, if $Q\prec S$ and $\lambda(Q)=0$. There could be several 'nested' resonances in the $S$ on the same or different level of embedding. We will call diagram (not obviously resonance) \textbf{irreducible}, if it haven't got 'nested' sub-diagrams.\\

Every reducible resonance diagram is a contraction of irreducible. Indeed, let $S=P\star Q$ be reducible resonance diagram, $\lambda(Q)=0$. Thus:
\begin{equation}
0=\lambda(S)=\lambda(P\star Q)=\lambda(P)+\lambda(Q)=\lambda(P) \Rightarrow \lambda(P)=0
\end{equation}
If $P$ - reducible, proposition is proved. If not, we could repeat this process. Such an algorithm of decomposition is finite, because $|P\star Q|=|P|+|Q|>|P|>0$.\\

In the vertex, where 'overlaped' resonances come to the new vertex, giving birth to a resonance, we will have an integral:
\begin{equation}
\int^{t_1}_{0}dt_2 t_2^{s}=\dfrac{t_1^{s+1}}{s+1}
\end{equation}

Here we have coefficient $1/(s+1)$, which breaks our 'superposition' principe for 'resonance lines'. Now, we are ready to write down a general algorithm for the resonance case. We subdivide it onto three parts. The main algorithm is:
\begin{enumerate}
\item First summation goes over all irreducible non-resonance (normalized) diagrams $\tilde{D}$, which could be constructed from the r.h.s. of equation.
\item In each term of the summations we have $\tilde{D}^i_{j_1...j_s}$.
\item Second summation goes over all possible ways of contraction of free bottom indexes of $\tilde{D}$ with $c^j$ - deformed initial conditions (see the third algorithm) or $\tilde{S}^{j_k}_k(s)$ - the resonance sub-diagrams coefficients (see the second algorithm), for all alowed $s$.
\item Multiply each term on the $e^{t\lambda_j}$ according to every $c^j$ contracted with bottom indexes of $\tilde{D}$.
\item Multiply each term by the $\left(t+L_{\tilde{D}}(v,i)\right)^s$, where $v$ - every vertex, where $\tilde{S}^{j_k}_k(s)$ is contracted with the bottom indexes of $\tilde{D}$; $i$ - initial vertex of $\tilde{D}$: such a vertex, that contains only upper index of $\tilde{D}$.
\end{enumerate}

On each free edge $v$ of $\tilde{D}$ we could collect all expressions contracted with it, and get:
\begin{equation}
X^i(\tilde{D},v,t)=c^i e^{t\lambda_i}+\sum\limits_{\tilde{S}}\sum\limits_{s=0}^{R(\tilde{S})} \tilde{S}^i(s) \left(t+L_{\tilde{D}}(v,i)\right)^s
\end{equation}
where $\sum_{\tilde{S}}$ - summation over all resonance diagrams, $R(\tilde{S})$ - number of irreducible resonance sub-diagrams of $\tilde{S}$. But, unfortunately, $X$ depends on the diagram $D$ and even on its free edge $v$, which it has been contracted with.\\

In the second algorithm, we calculate pure resonance contributions $\tilde{S}^{i_k}_k(s)$. For each resonance diagram $\tilde{S}^{i_k}_k$:
\begin{enumerate}
\item Write down $(\tilde{S}_k)^{i_k}(c)=(\tilde{S}_k)^{i_k}_{j_1...j_{s_k}}c^{j_1}...c^{j_{s_k}}$, where $c^{j_u}$ - the 'deformed' initial conditions.
\item Decompose $\tilde{S}$ into contraction of irreducible resonance diagrams $\tilde{S}_1,...,\tilde{S}_r$.
\item Summation goes over all possible scenarios of $(p-s)$ 'resonance lines' 'dying out', where $p$ - number of irreducible resonance diagrams in $\tilde{S}$.
\item Each scenario describing by choosing of $w_p$'s: each 'resonance line' getting its start in the initial vertex $v_p$ of some $\tilde{S}_p$'s and goes to the initial vertex of diagram $\tilde{D}$. For each 'resonance line', we need to choose  another vertex $w_p$ (which is also an initial vertex for some $\tilde{S}_q$) on the way from $v_p$ to $i$, for 'dying out'.
\item Multiply diagram by the $\prod_{q}L_{\tilde{S}}(v_q,w_q)$
\item Multiply each initial vertex of $\tilde{S}_p$'s by the $1/(l+1)$, where $l$ - the number of 'resonance lines', which passing this vertex.
\end{enumerate}

And in the third algorithm, we calculate 'deformed' initial conditions $c^i$:
\begin{enumerate}
\item First summation goes over all diagrams $\tilde{F}^i_{j_1...j_s}$, which could be constructed from the r.h.s. of equation.
\item Second summation goes over all partitions $\pi\in\mathcal{P}(F)$.
\item For each $D\in\pi$, write down normalized $\tilde{D}$.
\item Decompose $\tilde{D}$ into contraction $\tilde{D}^i=\tilde{Q}^i_{k_1}..._{k_a}..._{k_b}\tilde{S_1}^{k_1}...\tilde{S_a}^{k_a}..._{k_b}$, where $\tilde{Q}$ - irreducible non-resonance diagram, $\tilde{S}_{c}$ - resonance diagrams. Summate over all possible ways of choosing of $s_c$'s ($s_c$ have the same role as $s$ in the $\tilde{S}^{i_k}_k(s)$).
\item Multiply each $\tilde{S_c}$ by the coefficient, similar to the one, obtained in the previous algorithm, for some $s_c$.
\item Multiply each term by the $\prod_{v_c}\left(L_{\tilde{Q}}(v_c,i)\right)^{s_c}$, where $v_c$ - every vertex, where some $\tilde{S}_c$ is contracted with a bottom index of $\tilde{Q}$; $i$ - initial vertex of $\tilde{Q}$, $s_c$ - number, chosen for a diagram $\tilde{S}_c$, defined below.
\end{enumerate}

We can interpret each irreducible resonance diagram as a new effective degree of freedom of the system, dynamic along which is simple trivial $t$, and solution of the system in some neighborhood of the fixed point would look like:
\begin{equation}
x(t)=f(c^1 e^{t\lambda_1},...,c^d e^{t\lambda_d}, t,..., t)
\end{equation}
where each $t$ corresponds to some irreducible resonance diagram, $c^m$ - deformed initial conditions. Or, if we decompose the map:
\begin{equation}
f: \mathbb{C}^{d}\times \mathbb{C}^{r} \to \mathbb{C}^{d}
\end{equation}
where $r$ - number of irreducible resonance diagrams (not obviously finite number), dynamic in $\mathbb{C}^{d}$ would look like:
\begin{equation}
\dot{x}^i=\lambda^ix^i, x^i(0)=c^i
\end{equation}
And dynamic in $\mathbb{C}^{r}$ would look like:
\begin{equation}
\dot{y}^k=1, y^k(0)=0
\end{equation}
Resonance part of 'linearized' phase space definitely have finer structure.

\section{Vector fields algebra}

Vector fields space, which we have investigated, constitutes a Lie algebra relating to a commutation operation:
\begin{equation}
U=\sum u^i(x)\partial_i, V=\sum v^j(x)\partial_j
\end{equation}
\begin{equation}
\left[ U,V \right] =\sum (U^i\partial_i V^j-V^i\partial_i U^j)\partial_j
\end{equation}

Evolution operator is an action of some subgroup of a group $\mathrm{Diff}(\mathbb{C}^d)$, gotten by exponentiation, on $\mathbb{C}^d$. A problem of the formalism developed here for initial subgroup investigation, is that evolution operators action is defined locally, by series. It's not clear how to get global action properties on the whole $\mathbb{C}^d$. The global properties of the Lie group reconstruction by exponential elements is a complicated task.\\

The interesting properties of the considered algebras are that their 'universal enveloping' algebra has 'branching algebra' structure - composition of pairs of elements defined not uniquely, e.g.:
\begin{equation}
A^i_{j,k}\circ B^a_{b,c}=A^i_{j,k}B^j_{b,c}
\end{equation}
or
\begin{equation}
A^i_{j,k}\circ B^a_{b,c}=A^i_{j,k}B^k_{b,c}
\end{equation}

We could contract a pair of elements by different indexes, and get different result. The Lie algebra structure could be restored, by introducing a commutator:
\begin{equation}
[U,V]=\sum\limits_{k=0}^{a}U^i_{i_1...i_a}V^{i_k}_{i_{a+1}...i_{a+b}} - \sum\limits_{k=0}^{b} V^i_{i_1...i_b} U^{i_k}_{i_{b+1}...i_{a+b}}
\end{equation}
which we get from the vector fields commutator (or from quadratic part of evolution operators commutator). If we continue an analogy with linear algebra, we could introduce something like a linear operator product:
\begin{equation}
U\*V=\sum\limits_{k=0}^{a}U^i_{i_1...i_a}V^{i_k}_{i_{a+1}...i_{a+b}}
\end{equation}
and the operators commutator gets a traditional form:
\begin{equation}
[U,V]=UV-VU
\end{equation}
Unfortunately, this supporting product is not even associative (evolution operators' composition is, obviously, associative). We also could make an assumption, that according to this product, an evolution operator could be simply represented as:
\begin{equation}
x^i(t)=\mathrm{exp}(At)\circ x(0)=\sum\limits_{k=0}^{\infty}\dfrac{(At)^k}{k!}\circ x(0)
\end{equation}
A direct proof of this fact is related to some combinatorial difficulties.\\

In the same way, as we could consider the matrix Lie algebras in abstract way, forgetting about their matrix nature, coming back to it by representation - branching algebra is associative algebra, a product of pair of the elements in which could be done in the different ways. Representation of branching algebra should be done not to $\mathrm{Hom}(V,V)$, but to $\mathrm{Hom}(T(V),V)$:
\begin{equation}
R: A\to \mathrm{Hom}(V,T(V))
\end{equation}
where $T(V)=\bigoplus_{s=0}^{\infty}V^{\otimes s}$. If we act with $A$ on $V$, all interesting non-linearity is coming from a 'lifting to the diagonal' operator:
\begin{equation}
N: V \to T(V)
\end{equation}
$$
v \mapsto \bigoplus\limits_{s=0}^{\infty}v^{\otimes s}
$$
so the action of the algebra $A$ on $V$ is delivering by an composition:
\begin{equation}
R(A)\circ N: V \to V
\end{equation}

Some freedom in $R$ choosing also exists: it is easy to see, that exists $R_1 \neq R_2$ such that $\forall \alpha\in A: R_1(\alpha)\circ N= R_2(\alpha)\circ N$.\\

Actually, the considered structure has not linear, but rather affine nature. To illustrate this idea, let's consider an example.\\

\textbf{Example.} We consider sub-algebra $L$ of polynomial vector fields algebra, which has non-degenerated zeroes in the points $z_1,...,z_p$:
\begin{equation}
U=u^i\partial_i
\end{equation}
\begin{equation}
u^i(z_k)=0, \dfrac{\partial u^i}{\partial z^j}(z_k) \neq 0
\end{equation}

When we decompose it in the neighborhood of non-degenerated zeroes, it has the following form:
\begin{equation}
U=a^i_j z^j\partial_i + b^i(z)\partial_i=z A \partial + b(z)\partial
\end{equation}
\begin{equation}
V=c^i_j z^j\partial_i + d^i(z)\partial_i=z B \partial + d(z)\partial
\end{equation}
where $A$ and $B$ - some matrices, $b(z)$ and $d(z)$ - polynomials. The commutators of these fields would look like:
\begin{equation}
\left[ U,V \right]=z(AB-BA)\partial + f(z)\partial
\end{equation}
where $f(z)$ - some new polynomial. A linear part of the vector field in the neighborhood of every point commutes in the same way, independently of the chosen coordinates. Thus, if we consider an algebra of vector fields, in which to every zero $z_p$ some matrix Lie algebra $L_p$ is assigned, and linear part of field in the neighborhood of zeroes is fixed by corresponding algebra, then such algebra has to be closed according to the commutator. The group of algebra $L$ would 'contain' the groups of algebras $L_p$ 'inside'.

\section{Discussion}

The diagrammatic technique considered in the text allows us to look at a non-linear dynamic in continuous time from algebraic point of view. In fact, the key result is the formula (\ref{general_exponent}) for non-linear exponent:
\begin{equation}
x^i(t)=ev^i\{V|t\}x(0)=\sum\limits_{D}\dfrac{t^{|D|}}{D!}D^i(x(0))
\end{equation}
which develop a map from Lie algebra of the vector fields to a corresponding group.\\

Another key result is an algorithm of the resonance point neighborhood dynamic calculations, described in the section 'Perturbations theory: resonance case', and the idea of the arising of additional degrees of freedom as a consequence of the resonances.\\

A scale of the considered phenomenons, in some sense, is intermediate between local, in which systems behavior is determined by constant and linear terms, and global one, in which a global geometry of vector field, singularities, zeroes mutual disposition, etc. is sufficient.\\

In conclusion, let's consider some questions, which look worth to be discussed later:\\
\textbf{Applications:}
\begin{enumerate}
\item \textit{Chaos theory.} It is well known, that chaotic dynamic could be observed in spatial dimension 3 or higher only. In context of resonances in the perturbations theory, we could suppose, that it is connected to very large number of solutions of the equations $\sum\lambda_k n_k=0$ for particular $\lambda$ values. How bifurcations are related to the dependency of a number of solutions on parameters?
\item \textit{Hamiltonian dynamic.} Under such consideration, in Hamiltonian systems a non-linearity looks like $\sum_kS^{jj_k}T_{j_1...j_s}$, where $S^{jj_k}$ - symplectic form. An interesting question is what specific properties Hamiltonian dynamic has in this context, how this applies to KAM-theory?
\item \textit{Integrable systems.} The symmetries groups, which arise in integrable systems, are often generated by some semi-simple Lie algebras. We could suppose, that when the system leaves a regime, where system is linear up to deformation, a general group, generated by the arbitrary vector fields would play a substantial role. In its study, the examples, which arise from the polynomial vector fields could be useful.
\end{enumerate}
\textbf{Calculations:}
\begin{enumerate}
\item \textit{Fixed point}. How can we get the series near a fixed point not from the ordered integration, but from the general formula for evolution operator?
\item \textit{Resonance phase space structure.} How could we describe better topology of a phase space near the resonance fixed point?
\item \textit{Convergence.} How can we compute a convergence domain of an arbitrary evolution operator, around some point? And In perturbation serieses case?
\item \textit{Jordan form.} How should we change ansatz, to consider dynamic around the fixed points, with non-diagonisable linear part of field?
\item \textit{Realification.} All calculations in this text was done over field $\mathbb{C}$. When we do calculations over $\mathbb{R}$, some linear operators couldn't be completely diagonalized - some non-Jordan cells is left sometimes. What new phenomenons would it produce?
\item \textit{Diagrams reduction}. How could we simplify the calculations, in case if non-linearity could be represented as a product $P^i_{j_1...j_s}=Q^i_{j_1...j_r}S_{j_{r+1}...j_s}$?
\end{enumerate}
\textbf{Conceptual:}
\begin{enumerate}
\item \textit{Approximations.} If we approximate some function in the r.h.s of equation by polynomial of some degree, how an increase of order of this polynomial change topology of a phase space of the dynamical systems?
\item \textit{'Branching' algebras.} What is the difference of the branching algebras from the ordinal one? How their representations theory in tensor algebra looks like? What is the specific properties of corresponding Lie groups? 
\item \textit{Normal forms.} A linear operators classification is based on their normal form. It is also useful in the straightforward matrix exponent calculations. How could we classify the non-linear operators? This way, we should get better understanding of the resultants theory (see \cite{Gelfand},  \cite{Resultants}).
\item \textit{Infinite dimensional operators.} What results this technique could develop in case of PDE? In fact, all which we should change is to replace a finite dimensional vector space, with the Hilbert space of functions over some manifold. Another approach is a space discritization, and a continuum limit. For example, KdV equation is a bright example of decomposition around fixed point - it could be represented as a sum of a linear operator and a polynomial non-linearity.
\end{enumerate}

\appendix

\section{Resonance terms}

Resonance terms for non-linearity of the second degree, and corresponding frequencies relation:\\
\begin{enumerate}
\item[] For $k=3$:
\begin{enumerate}
\item $y\to xx + x\to xx$
\item $y\to yx + y\to xx $
\end{enumerate}
\item[] For $k=4$:
\begin{enumerate}
\item $y\to xx + x\to xx$ (2 resonance diagrams)
\item $y\to xy + x\to xx + y\to xx$ (2 resonance diagrams)
\item $y\to xy + y\to xx$
\item $y\to xx + x\to yx$
\item $y\to yy + y\to xx $
\end{enumerate}
\end{enumerate}
The resonance terms for non-linearity of the third degree, and corresponding frequencies relation:
\begin{enumerate}
\item[] For $k=3$:
\begin{enumerate}
\item $y\to xxx$
\end{enumerate}
\item[] For $k=4$, there is no resonances
\item[] For $k=5$:
\begin{enumerate}
\item $y\to xxx + x\to xxx$ 
\item $y\to yxx + y\to xxx$
\end{enumerate}
\item[] For $k=6$:
\begin{enumerate}
\item $y\to xxx + x\to xxx$
\item $y\to yxx + y\to xxx$
\end{enumerate}
\item[] For $k=7$:
\begin{enumerate}
\item $x\to xxx+y\to xxx$ (2 resonance diagrams)
\item $y\to xxx+x\to yxx$
\item $y\to xxx+x\to xxx+y\to yxx$ (2 resonance diagrams)
\item $y\to yxx+y\to xxx$
\end{enumerate}
\item[] For $k=8$, no resonances.
\end{enumerate}
A theoretical end numerical check for the resonances have been done separately.

\section*{Acknowledgments}

I am highly grateful to Valerii Dolotin and Alexey Morozov, who involved me into this research field, and for our conversation, which guided me through this work. Also I want to thank Yaroslav Gerasimenko, Adel Salakh and Pavlo Havrilenko for our fruitful discussions.\\

This work was supported by the joint Ukrainian-Russian SFFR-RFBR project  F53.2/028.

\end{document}